\documentclass[twocolumn]{aastex63}
\usepackage{color, mhchem, here}
\usepackage{bm}
\usepackage{lineno}
\nonumber
\received{}
\revised{}
\accepted{}
\submitjournal{ApJ}

\shorttitle{Protoplanetary disk dispersal}
\shortauthors{Komaki et al.}

\newcommand{\rg}{r_{\rm g}}
\newcommand{\cs}{c_{\rm s}}

\newcommand{\au}{{\rm \,au}}

\newcommand{\Msun}{{\rm \, M_\odot}}
\newcommand{\Lsun}{{\rm \, L_\odot}}
\newcommand{\cm}{{\rm \, cm}}
\newcommand{\gram}{{\rm \, g}}
\newcommand{\second}{{\rm \, s}}
\newcommand{\erg}{{\rm \, erg}}
\newcommand{\eV}{{\rm \, eV}}
\newcommand{\keV}{{\rm \, keV}}
\newcommand{\dd}{{\rm d}}
\newcommand{\Kelvin}{{\rm \,K}}
\newcommand{\Myr}{{\,\rm Myr}}

\newcommand{\figref}[1]{Figure~\ref{#1}}
\newcommand{\tabref}[1]{Table~\ref{#1}}
\newcommand{\eqnref}[1]{Eq.~\ref{#1}}
\newcommand{\secref}[1]{Section~\ref{#1}}

\begin{document}

\title{Simulations of Protoplanetary Disk Dispersal: Stellar Mass Dependence of the Disk Lifetime}

\author[0000-0002-9995-5223]{Ayano Komaki}
\affiliation{Department of Physics, The University of Tokyo, 7-3-1 Hongo, Bunkyo, Tokyo 113-0033, Japan}
\email{ayano.komaki@phys.s.u-tokyo.ac.jp}

\author{Shuhei Fukuhara}
\affiliation{Department of Multi-Disciplinary Sciences, Graduate School of Arts and Sciences, The University of Tokyo, 3-8-1 Komaba, Meguro, Tokyo 153-8902, Japan}

\author{Takeru K. Suzuki}
\affiliation{Department of Multi-Disciplinary Sciences, Graduate School of Arts and Sciences, The University of Tokyo, 3-8-1 Komaba, Meguro, Tokyo 153-8902, Japan}
\affiliation{Komaba Institute for Science, The University of Tokyo, 3-8-1 Komaba, Meguro, Tokyo 153-8902, Japan}
\affiliation{Department of Astronomy, Graduate School of Science, The University of Tokyo, 7-3-1 Hongo, Bunkyo, Tokyo 113-0033, Japan}

\author[0000-0001-7925-238X]{Naoki Yoshida}
\affiliation{Department of Physics, The University of
Tokyo, 7-3-1 Hongo, Bunkyo, Tokyo 113-0033, Japan}
\affiliation{Kavli Institute for the Physics and Mathematics of the Universe (WPI), UT Institute for Advanced Study, The University
of Tokyo, Kashiwa, Chiba 277-8583, Japan}
\affiliation{Research Center for the Early Universe (RESCEU), School of
Science, The University of Tokyo, 7-3-1 Hongo, Bunkyo, Tokyo 113-0033, Japan}



\begin{abstract}
Recent infrared and submillimeter observations suggest that the protoplanetary disk lifetime depends on the central stellar mass.
The disk dispersal is thought to be driven by viscous accretion, magneto-hydrodynamics (MHD) winds, and photoevaporation by the central star.
We perform a set of one-dimensional simulations of long-term disk evolution that include all the three processes. We vary the stellar mass in the range of 0.5--7$\Msun$, and study the mass dependence of the disk evolution.
We show that a significant fraction of the disk gas is lost by MHD winds in the early stage, but the later disk evolution is mainly governed by photoevaporation.
The disk radius decreases as photoevaporation clears out the gas in the outer disk efficiently.
The qualitative evolutionary trends of the disk mass are remarkably similar for the wide range of the central stellar mass we consider, and the time evolution of the disk mass can be well fitted by a simple function.
The dispersal time is approximately 
ten million years for low mass stars with weak mass dependence, 
but gets as short as two million years around a $7\Msun$ star.
In the latter case, a prominent inner hole is formed by the combined effect of accretion and MHD winds within about one million years.
The strength of the MHD wind and viscous accretion
controls the overall mass-loss rate, but does not
alter the dependence of the dispersal timescale on the
central stellar mass.
\end{abstract}



\section{Introduction} \label{sec:intro}
Recent observations discovered diverse planetary
architectures around various types of stars \citep{Fulton:2017,Zhang:2018}.
Planets are formed in protoplanetary disks (PPDs) out of disk materials
within some limited time, and thus 
the dynamical evolution of a PPD
affects significantly planet formation.
In particular, disk dispersal process and its characteristic timescale
are thought to be critically important for setting the scene for the formation of diverse planets.

Observations of several star forming regions 
suggest that PPDs have a lifetime of a few million years \citep[e.g.,][]{Haisch:2001, Meyer:2007, Hernandez:2007, Mamajek:2009, Bayo:2012, Ribas:2014}.
It is also found that the disk lifetime may depend on the central stellar mass \citep[e.g.,][]{Carpenter:2006,Lada:2006, Allers:2007, DahmHillenbrand:2007, KennedyKenyon:2009, Fang:2012, Yasui:2014, Ribas:2015}.
For example, \cite{Ribas:2015} 
sorted disks by the system age and the central stellar mass, 
and calculated the disk fraction of each sample.
Based on the finding that the disk fraction decreased as the stellar mass increased, 
they concluded that disks around high-mass stars had shorter lifetimes.

The three major disk dispersal mechanisms proposed so far are 
viscous accretion onto the central star, magnetohydrodynamic (MHD) winds, and photoevaporation.
In a PPD, efficient angular momentum transfer causes the gas of the inner disk to fall onto the star.
The accretion rate is characterised by $\alpha$ parameter \citep{Shakura:1973,Lindenbell:1974}, which can be estimated in several ways based on observations \citep{Calvet:2005,Fedele:2010,Mathews:2012}.
Interestingly, a positive correlation is suggested between the stellar mass and the accretion rate  \citep{Muzerolle:2003,Hartmann:2006,HerczegHillenbrand:2008,Fairlamb:2015,Hartmann:2016}.

The role of MHD disk winds in the evolution of PPDs has attracted considerable attention in recent years \citep[e.g.,][]{Pascucci:2022}. 
In addition to the direct mass loss \citep{Suzuki:2009}, MHD disk winds remove angular momentum from the disk, which promotes mass accretion \citep{Blandford:1982,Bai:2016,Bethune:2017,Gressel:2020}. 
As a result, the radial profile of surface density is significantly altered from the one without the effect of MHD disk winds \citep{Suzuki:2016,Hasegawa:2017,Tabone:2022}.
MHD-driven winds also cause a great impact on the evolution of solid particles in PPDs \citep{Taki:2021,Arakawa:2021} and the formation and migration of planets \citep{Ogihara:2018,Kimmig:2020}. 

Photoevaporation is driven by high-energy radiation such as far-ultraviolet (FUV; $6\eV \lesssim h\nu < 13.6\eV $), extreme-ultraviolet (EUV; $13.6\eV < h\nu \lesssim 100\eV$), and X-ray ($100\eV \lesssim h\nu \lesssim 10\keV$) photons, which
heat the gas on the disk surface to launch photoevaporative flows.
The effect of photoevaporation on the disk evolution has been studied by a number of authors \citep{Hollenbach:1994,RichlingYorke:1997,GortiHollenbach:2009,Ercolano:2009,Owen:2010,Tanaka:2013,WangGoodman:2017,Nakatani:2018a,Nakatani:2018b,Picogna:2019,Komaki:2021,Picogna:2021}.
\cite{GortiHollenbach:2009} performed 1+1 dimensional simulations of disk photoevaporation by 
varying the central stellar mass in the range of 0.3--7.0$\Msun$.
They incorporated both accretion and photoevaporation as major disk dispersal processes.
\cite{Komaki:2021} used
radiation hydrodynamics simulations to
show that the mass-loss rate by photoevaporation increased as the central stellar mass.

It is important to notice that disk dispersal may likely be caused by a combination of
multiple physical mechanisms, between which there can be complicated interplay, 
and that the whole dispersal process can last over a long time of several to ten million years.
Clearly, it is necessary to study the long-term disk evolution 
with incorporating all the proposed dispersal mechanisms in a consistent manner. 
This motivated several theoretical and numerical studies
\citep{Clarke:2001,Gorti:2015,Kunitomo:2020, Kunitomo:2021},
but none of them consider accurate radial {\it profiles} of photoevaporation that are derived from detailed radiation hydrodynamics calculation with non-equilibrium chemistry.

In the present paper, we perform a set of long-term one-dimensional simulations of PPD evolution considering realistic photoevaporation profiles.
Our calculations incorporate,
for the first time, physical models of viscous accretion, MHD winds, and photoevaporation for PPDs around a wide range of the central stellar mass.
We follow the disk evolution until the disk is dispersed nearly completely, and derive the disk lifetime accurately. We also study in detail the mass-loss processes at a variety of evolutionary phases.

The rest of the paper is organized as follows.
In Section 2, we explain the methods we apply.
In Section 3, we show the main results.
In Section 4, we discuss detailed properties of our model calculations.
Finally in Section 5, we summarise the paper.

\section{Numerical methods}
We perform long-term disk evolution simulations varying the central stellar mass in the range of 0.5--7.0$\Msun$.
We adopt the cylindrical coordinates $(r,\phi,z)$
assuming the disk is axisymmetric about the z-axis.
We follow the evolution of the disk surface density, $\Sigma=\int\rho\,\dd z$, which is the integrated gas density in $z$-direction.
We calculate the time evolution of the gas temperature consistently 
until the disk disperses.
We incorporate accretion, MHD winds and photoevaporation in the following manner.
The governing equations are
\[
\begin{split}
&\frac{\partial\Sigma}{\partial t} + \frac{1}{r}\frac{\partial}{\partial r}\left(r\Sigma v_{r}\right) + \dot{\Sigma}_{\textrm{w}} + \dot{\Sigma}_{\textrm{pe}}=0,\\
&r\Sigma v_{r} = -\frac{2}{r\Omega}\left[ \frac{\partial}{\partial r}\left(r^2\Sigma\overline{\alpha_{r\phi}}\cs^2\right)+r^2\overline{\alpha_{\phi z}}\left(\rho\cs^2\right)_{\textrm{mid}} \right],
\end{split}
\]
where $\dot{\Sigma}_{\textrm{w}}$ and $\dot{\Sigma}_{\textrm{pe}}$ are the surface mass-loss rates by MHD winds and photoevaporation, respectively.
Details of the mass-loss profiles by MHD winds and by photoevaporation are described later in this section.
We incorporate the two mass-loss mechanisms as a simple sum of  $\dot{\Sigma}_{\textrm{w}}$ and $\dot{\Sigma}_{\textrm{pe}}$.
Although this 
might overestimate the total mass-loss rates, we have confirmed that the disk evolution does not vary significantly if we use a conservative setup with taking the larger value of $\dot{\Sigma}_{\textrm{w}}$ and $\dot{\Sigma}_{\textrm{pe}}$, in which the only dominant mechanism operates at a given time.
This is simply because only one effect dominates for most of the time
during the disk evolution.
In the above equations, $v_{r}$, $\Omega$ and $\cs$ are the velocity in $r$-direction, the angular velocity and the sound speed respectively.
The subscript `mid' expresses the value at the mid-plane.
The dimensionless parameters $\alpha_{r\phi}$ and $\alpha_{\phi z}$ express the efficiency of the viscous and wind-driven accretion, respectively. The mass-weighted averages are denoted as $\overline{\alpha_{r\phi}}$ and $\overline{\alpha_{\phi z}}$ (see \cite{Suzuki:2016}
for the definition.)

We calculate the disk gas temperature by considering irradiation from the central star and viscous heating following \cite{Suzuki:2016}.
The $z$-averaged temperature is calculated from the balance equation
\[
T^4 = T_{\textrm{irr}}^4 + T_{\textrm{vis}}^4,
\]
where $T_{\textrm{irr}}$ and $T_{\textrm{vis}}$ are the temperatures corresponding to the energy by irradiation and by viscous heating.
The irradiation temperature is given by
\[
T_{\textrm{irr}} = 280\Kelvin \left(\frac{L_{*}}{\Lsun}\right)^{1/4}\left(\frac{r}{1\au}\right)^{-1/2},
\]
where $L_{*}$ is the stellar bolometric luminosity and $\Lsun$ is the solar bolometric luminosity.
Here we assume that the temperature reaches equilibrium quickly so that the gas temperature is equal to the dust temperature.
The above equation effectively approximates that the dust temperature is determined by the incident radiation from the central star.
The bolometric luminosity of the central star at the chracteristic age of $1\Myr$ is adopted from \cite{GortiHollenbach:2009}.
There are several stellar evolution models for pre-main sequence stars, but there remains a discrepancy especially for stars younger than $1\Myr$ \citep{Tout:1999,Siess:2000}.
For completeness of our model, we use the stellar luminosity at the age of $1\Myr$ (\tabref{tab:stellarparameter}).

The viscous heating temperature is determined by the equation
\[
2\sigma_{\textrm{SB}}T_{\textrm{vis}}^4=\left(\frac{3}{8}\tau_{\textrm{R}}+\frac{1}{2\tau_{\textrm{P}}}\right)F_{\textrm{rad}},
\]
where $\sigma_{\textrm{SB}}, \tau_{\textrm{R}}, \tau_{\textrm{P}}$, and $F_{\textrm{rad}}$ are Stefan-Boltzman constant, the Rosseland mean optical depth, the Planck mean optical depth, and the radiation flux from the disk surface \citep{NakamotoNakagawa:1994}.
This equation expresses the energy transfer in $z$-direction under the assumption of thermodynamical equilibrium in a geometrically thin disk.
The disk gas is heated by viscous dissipation, and the deposited thermal energy diffuses out to the disk surface. A part of the energy liberated by accretion is finally released from the disk surface as radiation.
The Rosseland mean optical depth is expressed as 
\[
\tau_{\textrm{R}}=\kappa_{\textrm{R}}\Sigma, 
\]
where $\kappa_{\textrm{R}}$ is the opacity of dust grains and given by \cite{HuesoGuillot:2005} as
\[
\kappa_{\textrm{R}}=
\left\{
\begin{array}{ll}
     4.5\left(\frac{T}{150\Kelvin}\right)^2 \cm^2\gram^{-1} & (T<150\Kelvin)\\
     4.5\cm^2\gram^{-1} & (150\Kelvin\leq T\leq 1500\Kelvin)\\
     0\cm^2\gram^{-1} & (T>1500\Kelvin)
\end{array},
\right.
\]
assuming silicate and water ice as the dominant opacity agents \citep{Pollack:1985}.
Note that the temperature dependence changes at around $150\Kelvin$ because of water ice depletion.
In our calculation, we adopt a smooth function 
\[
\begin{split}
\kappa_{\textrm{R}}=2.25\cm^2\gram^{-1}\,&\left[1-\tanh\left(\frac{T-1500\Kelvin}{150\Kelvin}\right)\right]\\
&\times\textrm{min}
\left[1, \left(\frac{T}{150\Kelvin}\right)^2\right]
\end{split}
\]
which approximates the above values well.
For low-temperature dust, the Rosseland and Planck mean opacities can be expressed by a power-law, 
and the relation $\tau_{\textrm{P}}=2.4\tau_{\textrm{R}}$
holds at low temperature \citep{NakamotoNakagawa:1994}.
We thus set the Planck mean optical depth as
\[
\tau_{\textrm{P}}=\textrm{max}(2.4\tau_{\textrm{R}}, 0.5),
\]
where a lower limit of 0.5 is adopted
to reproduce the optically thin limit.


We consider a wide range of the effective viscosity $\alpha$ as a parameter.
Recent ALMA observations toward star-forming regions suggest that the $\alpha$ parameter varies over two orders of magnitude \citep{Rafikov:2017,Ansdell:2018}.
\cite{Ansdell:2018} estimated the gas radius of each disk from $^{12}\ce{CO}$ line emission.
They simulated the evolution of gas radius with varying $\alpha$ parameter in the range of $10^{-4}$--$10^{-2}$ assuming an initial disk model by \cite{Facchini:2017} and compared the result with the observations.
They found a wide distribution of the value of $\alpha$ over a few orders of magnitudes.
Several observations have been conducted to estimate the disk accretion rate using the H$\alpha$ equivalent width \citep{Fedele:2010,Mathews:2012,Hartmann:2016}.
These studies also have shown that there is a large variation in the accretion rate.
The optical-UV observations toward T Tauri stars have shown that there is a positive relationship between the accretion rate and the stellar mass given as $\dot{M}_{\textrm{acc}}\propto M_{*}^2$ \citep{Muzerolle:2003,Calvet:2004,Hartmann:2006}.
We thus assume 
\[
\overline{\alpha_{r\phi}}\propto M_{*}
\]
to match the observed trend.
In general, viscous accretion is effective only in the radial portions of the disk
where magnetorotational instability (MRI hereafter) \citep{Velikhov:1959,Chandrasekhar:1961,BalbusHawley:1991} operates to generate strong turbulence.
While MRI actively induces MHD turbulence in the inner region $< 1\au$ and the regions near the disk surfaces owing to sufficient ionization, an MRI-inactive region called a dead zone \citep[e.g.,][]{Gammie:1996} is supposed to occupy $1 \lesssim r \lesssim $ a few tens $\au$ of a PPD because of insufficient ionization near the midplane \citep{MoriOkuzumi:2016, Pinte:2016, Flaherty:2017}.
On the other hand, a low but finite value of $\alpha_{r\phi}$ can still be sustained by purely hydrodynamical processes such as vertical shear instability \citep{UrpinBrandenburg:1998, Nelson:2013, LinYoudin:2015, Flock:2020, Manger:2020}.
Although the ionization degree is expected to be varied over a large distance from the central star, the detailed radial extent of the dead zone is not still well understood quantitatively. 
Considering viscous accretion is dominant in the inner region, we assume a constant $\overline{\alpha_{r\phi}}$ throughout the disk.
We adopt $\overline{\alpha_{r\phi}}=1.0\times10^{-4}(M_{*}/1\Msun)$ as our fiducial value. 
In order to examine the impact of enhanced accretion in MRI-active disks, we also perform simulations with $\overline{\alpha_{r\phi}}=1.0\times10^{-2}(M_{*}/1\Msun), 1.0\times10^{-3}(M_{*}/1\Msun)$.

We parametrise the wind loss term $\dot{\Sigma}_{\textrm{w}}$, following \cite{2010Suzuki}:
\begin{equation}
\dot{\Sigma}_{\textrm{w}} = (\rho\cs)_{\textrm{mid}}C_{\textrm{w}}=\frac{\Sigma\Omega}{\sqrt{2\pi}}C_{\textrm{w}},
\label{eq:MHDwinds}
\end{equation}
where $C_{\textrm{w}}$ expresses a dimensionless mass flux given by
\[
C_{\textrm{w}}=\textrm{min}\left(C_{\textrm{w},0},C_{\textrm{w},e}\right).
\]
Here, the constant maximum value $C_{\textrm{w},0}$ is estimated from local shearing box MHD simulations of \cite{Suzuki:2009}.
We adopt $C_{\textrm{w},0}=2.0\times10^{-5}$ for the MRI-active case and $C_{\textrm{w},0}=1.0\times10^{-5}$ for the MRI-inactive case.
We calculate the energetics-constrained mass flux following \cite{Suzuki:2016}.
We consider two cases with strong and weak winds.
In the strong wind case, all the liberated gravitational energy is used to drive MHD winds.
In this case, the mass flux $C_{\textrm{w},e}$ and the energy flux $F_{\textrm{rad}}$ are given by
\[
\begin{split}
    C_{\textrm{w},e} &=\textrm{max}\left[\frac{2}{r^3\Omega(\rho\cs)_{\textrm{mid}}}\frac{\partial}{\partial r}\left(r^2\Sigma\overline{\alpha_{r\phi}}\cs^2\right) + \frac{2\cs}{r\Omega}\overline{\alpha_{\phi z}}, 0\right] \\
    F_{\textrm{rad}} &= \textrm{max}\left[-\frac{1}{r}\frac{\partial}{\partial r}\left(r^2\Sigma\Omega\overline{\alpha_{r\phi}}\cs^2\right), 0\right].
\end{split}
\]
In the weak wind case, a relatively small fraction of the sum of the liberated gravitational energy and the energy by viscous heating is spent to drive MHD winds, and the rest is emitted as radiation.
We define the fractional ratio of energy used to launch winds as $\epsilon_{\textrm{rad}}$. 
We then set
\[
\begin{split}
    C_{\textrm{w}} &= (1-\epsilon_{\textrm{rad}})\left[ \frac{3\sqrt{2\pi}\cs^2}{r^2\Omega^2}\overline{\alpha_{r\phi}}+\frac{2\cs}{r\Omega}\overline{\alpha_{\phi z}} \right]\\
    F_{\textrm{rad}} &= \epsilon_{\textrm{rad}}\left[ \frac{3\sqrt{2\pi}(\rho\cs^3)_{\textrm{mid}}}{2}\overline{\alpha_{r\phi}}+r\Omega\overline{\alpha_{\phi z}}(\rho\cs^2)_{\textrm{mid}} \right].
\end{split}
\]
For the main results presented in the following, we adopt the strong disk wind case as our fiducial model. We discuss the effect of weak wind separately in Section 4, where we set $\epsilon_{\textrm{rad}}=0.9$ for the weak wind case.
Note that this is a limiting case in which the radiation loss is
maximally evaluated.

Local shearing-box simulations show that the MHD wind torque satisfies $\overline{\alpha_{\phi z}}\sim10^{-5}$--$10^{-3}$ and has a positive dependence on the strength of the net vertical magnetic field $B_{z}$ \citep{Bai:2013}.
The dependence is well approximated as
\[
\overline{\alpha_{\phi z}}\propto \left(\frac{B_{z}^2}{8\pi (\rho\cs^2)_{\textrm{mid}}}\right)^{0.66}.
\]
The variation of $\overline{\alpha_{\phi z}}$ can be described by the ratio of the surface density with respect to the initial value 
on the assumption that the vertical magnetic flux stays 
constant during the disk evolution.
Then $\overline{\alpha_{\phi z}}$ is expressed as
\begin{equation}
\overline{\alpha_{\phi z}} = \textrm{min}\left[10^{-5}\left(\frac{\Sigma}{\Sigma_{\textrm{int}}}\right)^{-0.66}, 1\right],
\label{eq:alpha-phiz}
\end{equation}
where $\Sigma_{\textrm{int}}$ is the initial surface density.
Since the evolution of the magnetic field
is still poorly understood, we place a conservative upper limit on the value of $\overline{\alpha_{\phi z}}$.
We should note that \cite{Armitage:2013} reported that the vertical magnetic field also diffuses away as the surface density decreases; in this case, the dependence of $\alpha_{\phi,z}$ on $\Sigma$ would be weaker than in \eqnref{eq:alpha-phiz}.

We adopt the profile of $\dot{\Sigma}_{\textrm{pe}}$ based on the calculations by \cite{Komaki:2021}.
In practice, we fit the results of the radiation hydrodynamics simulations by a function that combines a quadratic function and a function with negative power.
The radiation hydrodynamics simulations incorporate EUV, FUV and X-ray radiation from the central star.
\cite{Komaki:2021} show that strong FUV radiation heats the disk gas effectively and drives rapid mass loss around a high mass star. It is also found that the gas in the outer disk is heated to generate flared structure.

Previous studies have shown that the mass-loss rate by photoevaporation, $\dot{M}_{\textrm{pe}}$, does {\it not} strongly depend on the initial disk mass by performing hydrodynamics simulations \citep{Wolfer:2019, Nakatani:2021}.
$\dot{M}_{\textrm{pe}}$ is primarily determined by the density at the disk surface rather than the density at the mid-plane, which is directly connected to the total mass. 
For the same reason, we also assume that $\dot{\Sigma}_{\textrm{pe}}$ is constant throughout the calculation.

The initial surface density is configured following the minimum solar disk model \citep{Hayashi:1981}, which is defined as 
\[
\Sigma_{\textrm{int}}=\Sigma_{1\au}\left(\frac{r}{1\au}\right)^{-3/2}\exp{\left(-\frac{r}{r_{\textrm{cut}}}\right)},
\]
where $r_{\textrm{cut}}$ is a cut-off radius.
Motivated by the observation that the disk radius increases proportionally to the stellar mass \citep{Andrews:2018}, we assume $r_{\textrm{cut}}\propto M_{*}$.
We list the physical parameters of our simulations in \tabref{tab:stellarparameter}.
Note that we normalize the disk gas density (and hence mass) by $\Sigma_{\textrm{1\au}}$, the surface density at $r=1\au$.
We determine the value so that the total disk mass satisfies $M_{\textrm{disk}}=0.117(M_{*}/1\Msun)\Msun$ as \cite{Hayashi:1981} suggests.
\begin{table*}
  \caption{Fiducial stellar parameters in the model (adopted from \citet{GortiHollenbach:2009})}
  \centering
  \begin{tabular}{cccccccc} \hline
    $M_{*}$ ($\Msun$) & $L_{\textrm{bol}}$ ($\Lsun$) & $M_{\textrm{disk}}$ ($\Msun$) & $r_{\textrm{cut}}$ ($\au$) & log$L_{\textrm{FUV}}$ ($\erg\second^{-1}$) & log$\phi_{\textrm{EUV}}$ ($\second^{-1}$) & log$L_{\textrm{X-ray}}$ ($\erg\second^{-1}$)  \\ \hline 
    0.5 & 0.93 & 0.0585 & 15 & 30.9 & 40.1 & 29.8 \\
    0.7 & 1.72
    & 0.0819 & 21 & 31.3 & 40.5 & 30.2\\
    1.0 & 2.34 
    & 0.117 & 30 & 31.7 & 40.7 & 30.4 \\
    1.7 & 5.00 
    & 0.199 & 51 & 32.3 & 41.0 & 30.7 \\
    3.0 & 14.85 
    & 0.351 & 90 & 32.9 & 39.0 & 28.7 \\
    7.0 & 1687 
    & 0.819 & 210 & 36.5 & 44.1 & 30.8 \\ \hline
  \end{tabular}
  \label{tab:stellarparameter}
\end{table*}
We set the computational domain at $r_{\textrm{in}}=10^{-2}\au<r<r_{\textrm{out}}=10^{4}\au$.
We impose the zero-torque condition on the inner and outer boundaries \citep{Suzuki:2016, Kunitomo:2020}. 
We calculate the evolution until the disk mass decreases to
an extremely small value of $M_{\textrm{disk}}<10^{-10}\Msun$.
Since the disk mass decreases rapidly in the last stage, the calculation time does not change even if a different threshold mass is adopted, provided that it is $M_{\textrm{disk}}<10^{-5}\Msun$.

\section{Results}   \label{sec:results}
\subsection{Around a solar-mass star}
\begin{figure}
    \includegraphics[width=\linewidth,clip]{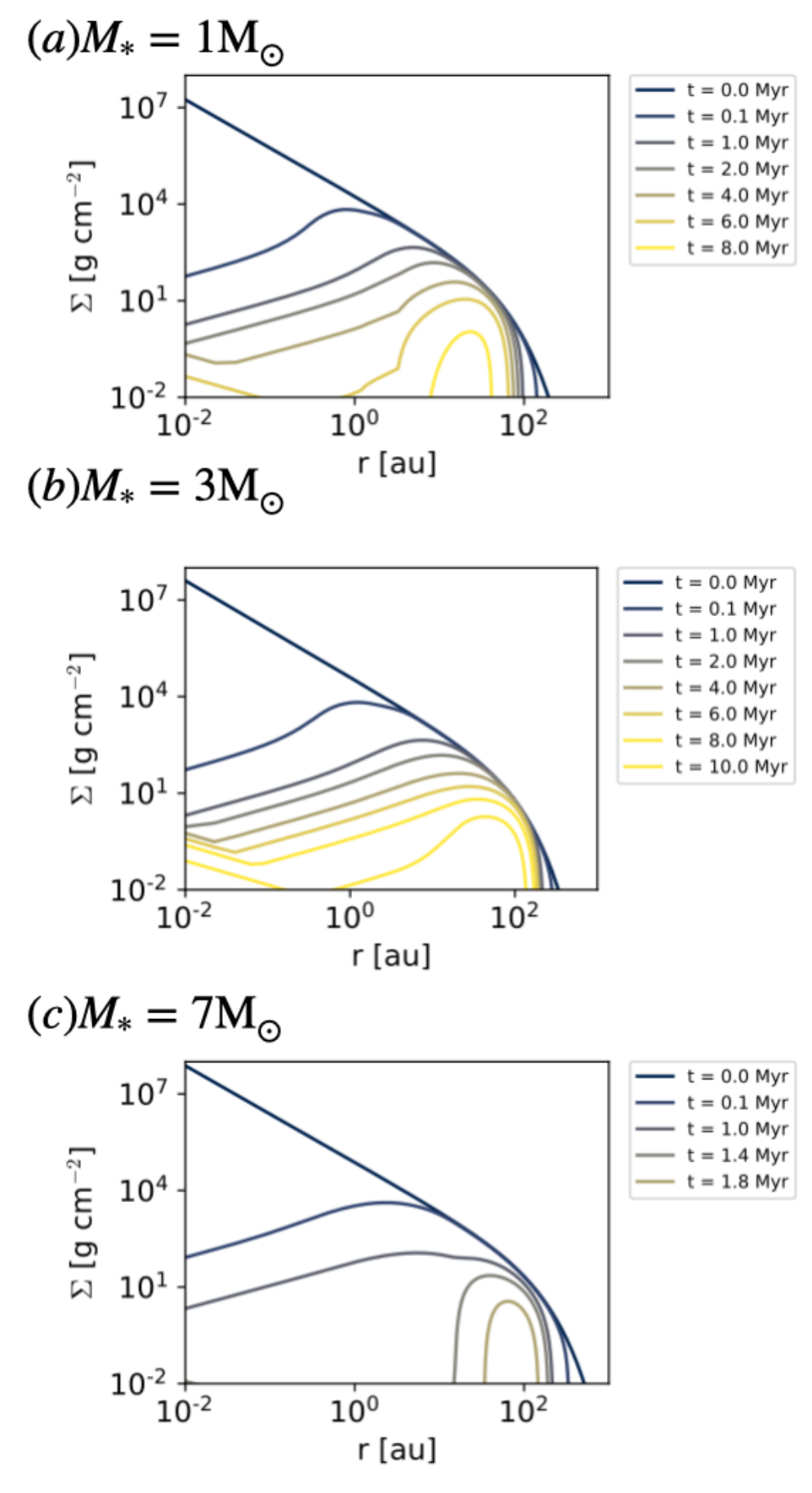}
    \caption{
    Snapshots of surface density in simulations with $M_{*}=1,3,7\Msun$.
    The initial surface density is shown in dark blue and as time goes on, the surface density is shown in more yellowish line.
    }
    \label{fig:snapshots}
\end{figure}
\figref{fig:snapshots} shows the snapshots from 0 to $8\Myr$ for the run with $M_* =1,3,7\Msun$.
In this section, we focus on the $M_* =1\Msun$ case and we explain the other cases in \secref{sec:3-2}.
The inner disk in the vicinity of the central star loses quickly a significant amount of mass by MHD winds in a few million years.
After $2\Myr$, the mass-loss rate by MHD winds, $\dot{M}_{\textrm{w}}$, decreases, and  
a steep density profile is found near the central star.
This is partly a numerical effect caused by the upper limit of $\overline{\alpha_{\phi z}}$ defined in \eqnref{eq:alpha-phiz};
the mass-loss rate by MHD winds is effectively limited. To examine the numerical effect,
we perform a set of simulations varying the upper limit of $\overline{\alpha_{\phi z}}$ by a factor of 0.1 and 10. We have confirmed that the mass-loss rate and the disk lifetime are not affected by this change.

\begin{figure}
    \includegraphics[width=\linewidth,clip]{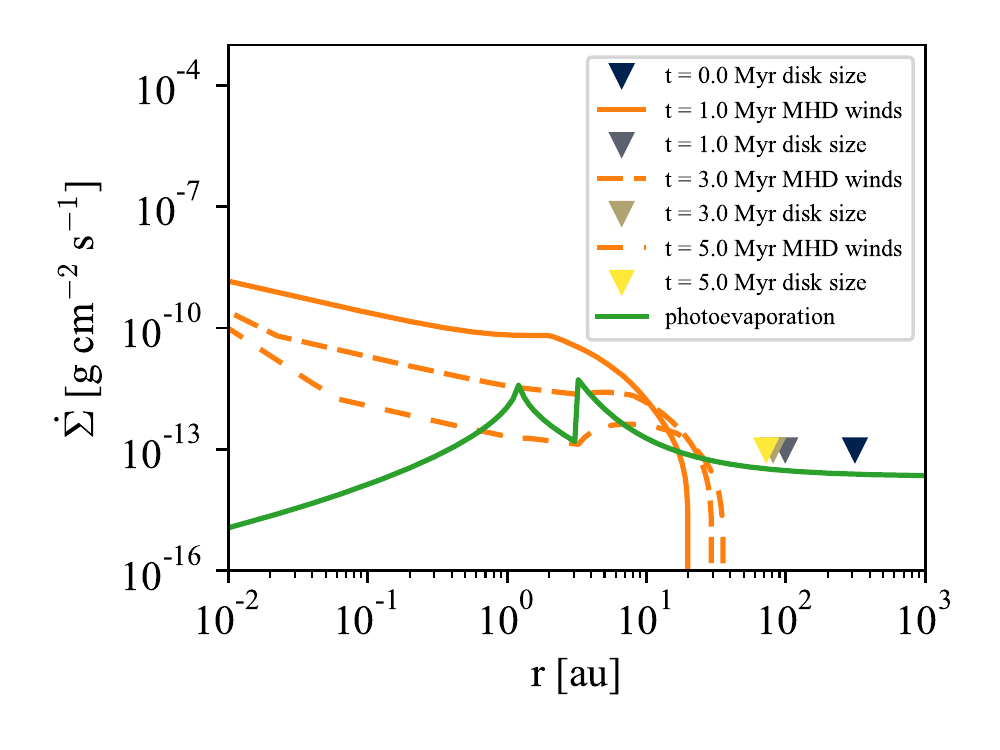}
    \caption{
    The surface mass-loss profiles by the two disk dispersal processes;  MHD winds(orange), photoevaporation(green) in the case of $M_{*}=1\Msun$.
    The triangles express the disk radius at the age of $0,1,3,5\Myr$.
    }
    \label{fig:dispersalsurfacedensity}
\end{figure}
\figref{fig:dispersalsurfacedensity} shows $\dot{\Sigma}_{\textrm{w}}$ and $\dot{\Sigma}_{\textrm{pe}}$ at the age of $t=1,3$ and $5\Myr$.
We also indicate by triangles the disk radii measured
from our simulation outputs.
Photoevaporation is the main dispersal process in the outer disk with $r\gtrsim20\au$ from the initial stage, 
while the inner disk is mainly dispersed by MHD winds.
As shown in \figref{fig:dispersalsurfacedensity}, $\dot{\Sigma}_{\textrm{w}}$ decreases with time because it is proportional to the surface density as given in \eqnref{eq:MHDwinds}.
These features are consistent with the results of \cite{Kunitomo:2020}.

An important result is that the disk radius decreases with time as indicated by the triangles in \figref{fig:dispersalsurfacedensity}.
Photoevaporation is the dominant dispersal process in the outer region, which we expect to cause the disk radius to decrease.
We have run test simulations without photoevaporation to study its effect clearly.
We have found that viscous accretion re-distributes the angular momentum and can indeed cause {\it bloating} of the disk even if $\overline{\alpha_{r\phi}}$ is set to a very low value of $10^{-4}$.
We thus conclude that photoevaporation plays a crucial role in shaping the disk morphology and its size through the effective mass-loss in the outer region.

 
At the age of $\sim 5 \Myr$, the major mass-loss is caused by photoevaporation in $r > 1\au$.
We explain the details of $\dot{\Sigma}_{\textrm{pe}}$ and investigate the impact of distinct photoevaporation profile on disk evolution later in \secref{sec:4-3}.

\begin{figure}
    \includegraphics[width=\linewidth,clip]{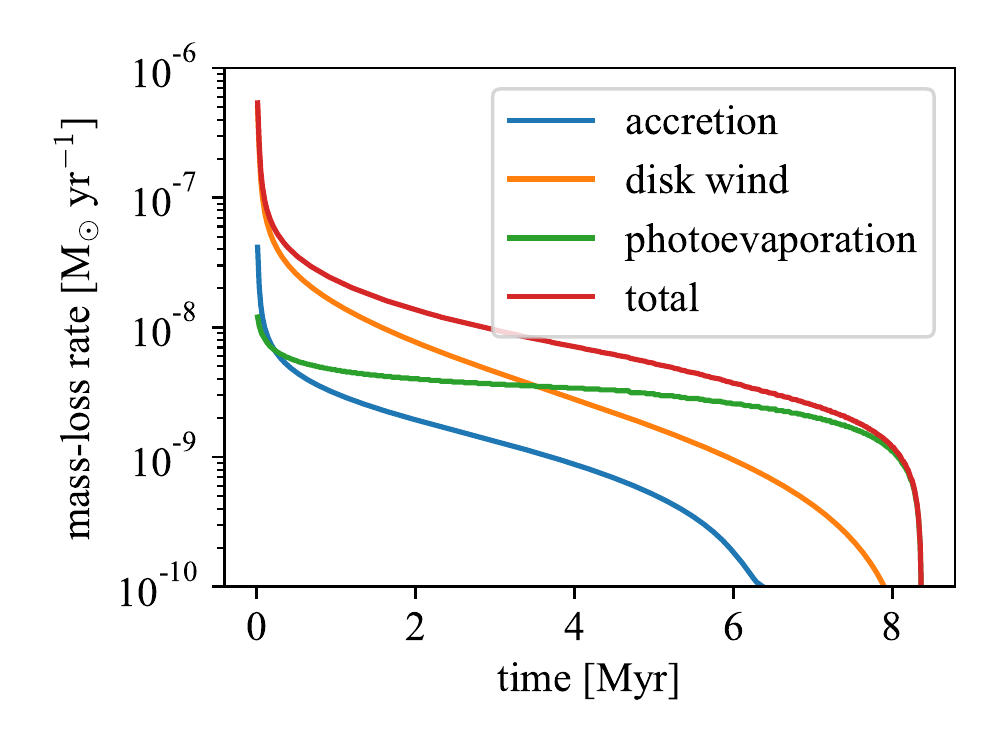}
    \caption{
    The mass-loss rates by the three dispersal processes; accretion (blue), MHD winds (orange), and photoevaporation (green)
    in our fiducial run with 
    $M_* = 1 \Msun$.
    The instantaneous mass-loss rates are calculated using \eqnref{eq:accretionprocess} and \eqnref{eq:dispersalprocess}.
    }
    \label{fig:dispersalprocess}
\end{figure}
\figref{fig:dispersalprocess} shows the mass-loss rate as a function of time.
The mass-loss rate by accretion is calculated as
\begin{equation}
\begin{split}
& \dot{M}_{\textrm{acc}}=-2\pi(r v_r \Sigma),\\
\label{eq:accretionprocess}
\end{split}
\end{equation}
using the values at the inner boundary.
The mass-loss rates by MHD winds and photoevaporation are calculated as follows.
\begin{equation}
\begin{split}
& \dot{M}_{\textrm{w}}=\int_{r_{\textrm{in}}}^{r_{\textrm{out}}^{\prime}} \dot{\Sigma}_{\textrm{w}}\,\dd r\\
& \dot{M}_{\textrm{pe}}=\int_{r_{\textrm{in}}}^{r_{\textrm{out}}^{\prime}} \dot{\Sigma}_{\textrm{pe}}\,\dd r,\\
\end{split}
\label{eq:dispersalprocess}
\end{equation}
where $r_{\textrm{out}}^{\prime}$ is the disk radius.
Initially, strong MHD winds disperse a large amount of mass, and photoevaporation becomes the dominant dispersal process after a few Myrs,
as has been discussed in the above.
In contrast, $\dot{M}_{\textrm{pe}}$ is almost constant with time.
The slight decrease in the photoevaporative mass-loss (\figref{fig:dispersalprocess}) is explained by the decreasing outer radius of the disk.


We integrate the mass-loss rate over time to derive the total mass lost.
We calculate the fractional contributions from the three
dispersal processes. The fraction of accretion, MHD winds and photoevaporation are $24\%$, $62\%$ and $13\%$,
respectively.
Clearly, more than a half of the disk mass is dispersed by MHD winds.
Since $\dot{\Sigma}_{\textrm{w}}$ scales proportional to the surface density,
the winds are blown with high densities from near the central star in the initial phase.
We note that previous studies suggest that $\dot{M}_{\textrm{w}}$ may often be overestimated. 
The 3D MHD simulations of an accretion disk show that a part of the gas launched as MHD winds falls back onto the central star through funnel-walls \citep{Takasao:2018,Takasao:2020}. 
In practice, the partially re-accreted gas should be included in $\dot{M}_{\textrm{acc}}$ rather than in $\dot{M}_{\textrm{w}}$.
For a similar reason, the upper limit of $C_{\textrm{w}}$ should be appropriately 
calibrated in our future work.

Protostellar jets can be 
another mass dispersal process. 
The mass-loss rate due to jets is estimated to be about $10\%$ of the accreted mass around low-mass stars \citep{Calvet:2004,Natta:2014}.
Since accretion itself is a relatively minor mass-loss process 
according to our calculations, we expect that the influence of jets on the disk dispersal would be limited.


\subsection{Simulations with different stellar masses}
\label{sec:3-2}
We perform disk evolution simulations with varying the stellar mass in the range of 0.5--7$\Msun$.
We compute the mass-loss rate by each process in the same manner as in the 1$\Msun$ case.
In all cases, the disk is dispersed dominantly by MHD winds in the early phase, and then photoevaporation replaces to become the dominant disk dispersal process in the last stage.

We find very similar disk evolution in the runs with lower mass stars ($< 1\Msun$).
Initially, MHD winds disperse the disk gas, and
photoevaporation becomes the main dispersal process at  $t \sim 4$--$5\Myr$.
Similarly to our $1\Msun$ run, 
about $60\%$ of the total mass-loss is due to MHD winds.

The case with $3\Msun$ shows a noticeable difference.
The evolution of the surface density is shown in the middle panel of \figref{fig:snapshots}.
The main disk dispersal process is MHD winds and later switches to photoevaporation as in our $1\Msun$ run,
but the transition occurs later at the age of $\sim6\Myr$.
Around a $3\Msun$ star, $\dot{M}_{\textrm{pe}}$ is relatively low,
because an intermediate-mass star does not have a well developed convective layer and the magnetic activity on the stellar surface is weak and hence generates lower X-ray emission.
The low X-ray luminosity results in the overall low mass-loss rate by photoevaporation \citep{Komaki:2021}.

The run with $M_{*}=7\Msun$ shows another difference; the disk is dispersed efficiently by both MHD winds and photoevaporation.
While MHD winds dominate the mass-loss in the very initial phase, the primary mechanism shifts to photoevaporation already at $t=1\Myr$ because the inner disk is completely dispersed as shown in \figref{fig:snapshots}.
The contribution of accretion, MHD winds and photoevaporation to the total mass loss is $23\%$, $45\%$ and $32\%$, respectively.
The relative contribution from the photoevaporation is larger than the other lower mass cases.

\begin{figure*}
    \includegraphics[width=\linewidth,clip]{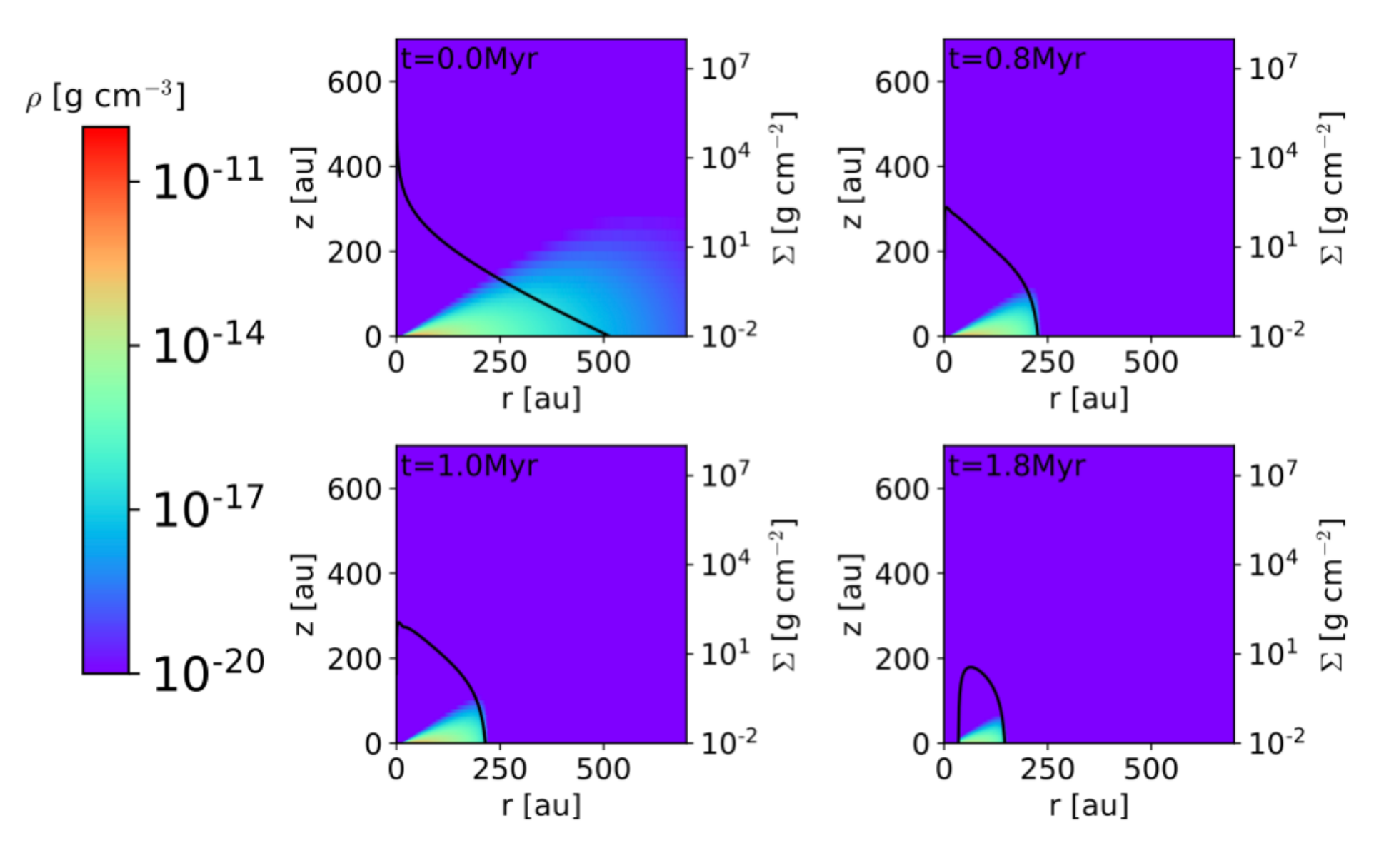}
    \caption{
    Snapshots of the reconstructed gas density in a simulation with $M_{*}=7.0\Msun$ at $t=0, 0.8, 1.0, 1.8\Myr$.
    The color map shows the density and the black line expresses the surface density.
    }
    \label{fig:snapshots-highmass}
\end{figure*}

The disk structure and its evolution can be seen 
more clearly by performing the following visualization.
We reconstruct two-dimensional density maps using the surface density obtained in the simulation with $M_{*}=7\Msun$.
\figref{fig:snapshots-highmass} shows the snapshots of the reconstructed disk density maps.
The vertical scale height throughout the disk is calculated by assuming hydrostatic equilibrium in the $z$-direction.
We see clearly that the diffuse gas in the outer disk disperses quickly in less than 1 Myrs. Photoevaporation critically affects the overall disk shape even though
its contribution to the total mass loss is small 
in the early epoch.
The inner disk has a small scale height, and the gas density is high on and near the mid-plane.
An inner "hole" opens at $\sim10\au$ at $t  \sim1.4\Myr$ by the combined effect of accretion and MHD winds, but the disk gas remains at $10\au\lesssim r \lesssim$ 200 $\au$ where the effect is relatively weak.


\subsection{stellar mass dependence of the disk lifetime}
\begin{figure}
    \includegraphics[width=\linewidth,clip]{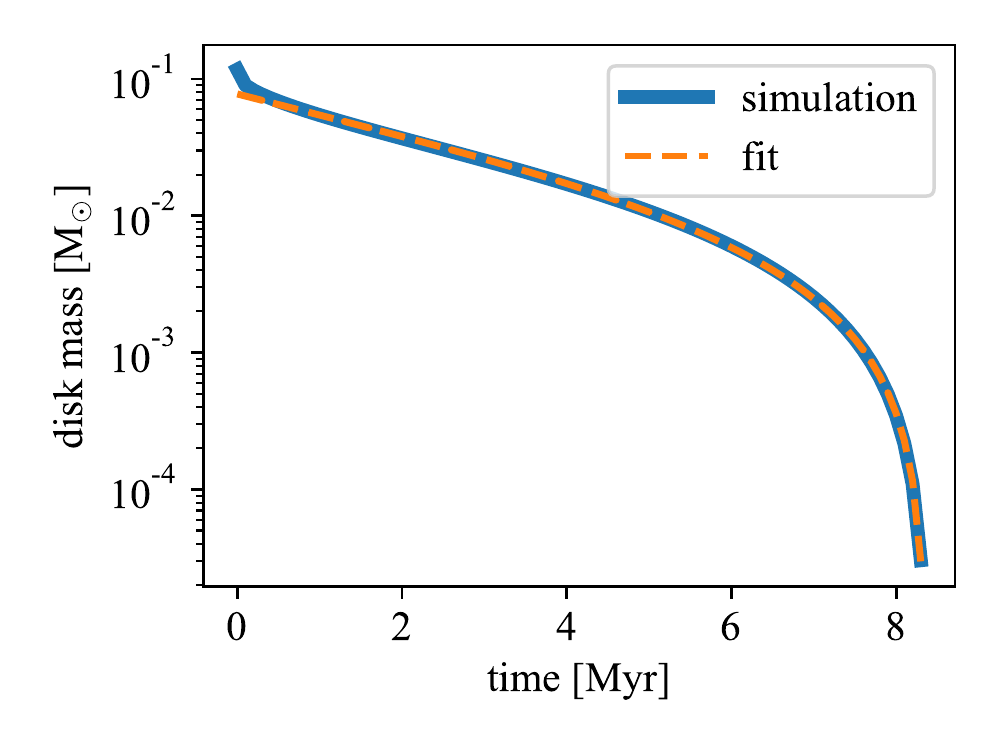}
    \caption{
    The disk mass evolution around a $1\Msun$ star.
    The blue line shows the simulation data and the orange dotted line expresses the fit.
    }
    \label{fig:diskmass}
\end{figure}
Based on our simulations, we propose an objective way of determining the disk lifetime.
\figref{fig:diskmass} shows the evolution of the disk mass in the case of $M_{*}=1\Msun$.
We find that the time evolution is accurately fitted with a simple vertical sigmoid function given by
\[
\begin{split}
\log_{10}M_{\textrm{disk}} &= a\log\left(\frac{1-x}{x}\right)+d\\
x &= \frac{(t/1\Myr)-b}{c}.
\end{split}
\]
We treat $a, b, c, d$ in the equation as fitting parameters.
With this functional form, the characteristic dispersal time is given by $(b+c)$ Myr.
\begin{figure}
    \includegraphics[width=\linewidth,clip]{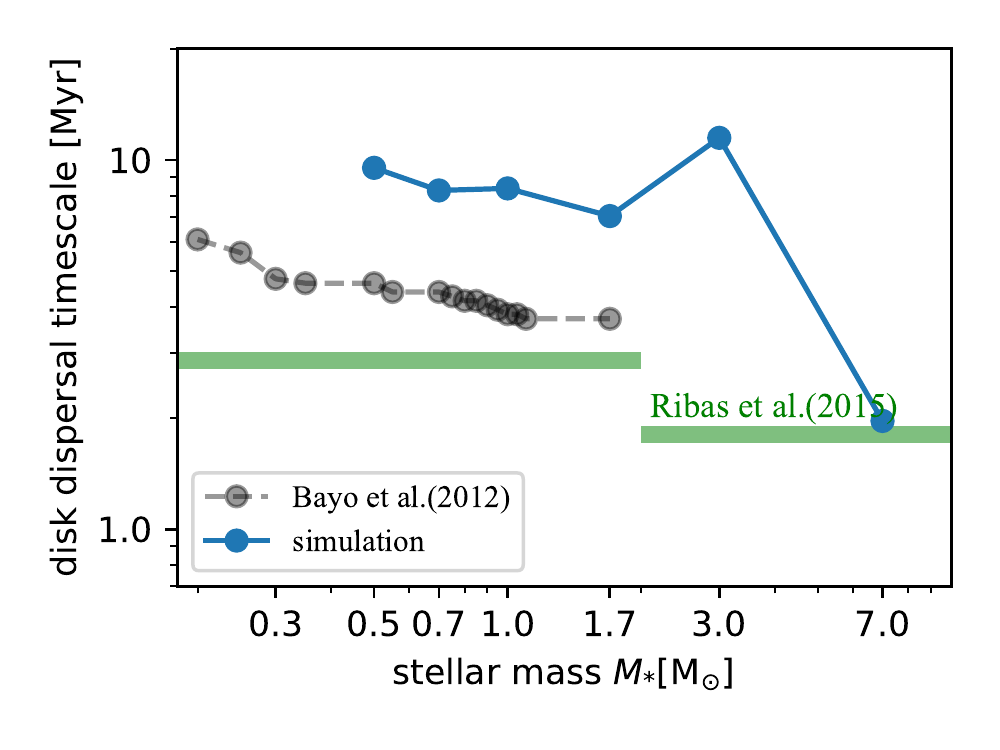}
    \caption{
    The disk dispersal timescale of each stellar mass case.
    The blue circles are the simulation data.
    The black points and green horizontal bars represent inner-disk lifetimes estimated from observational disk fractions in \cite{Bayo:2012} and \cite{Ribas:2015}, respectively. See the main text for more detail. 
    }
    \label{fig:disklifetime}
\end{figure}
We do the fitting and estimate the disk 
dispersal times for all the cases we simulated. In \figref{fig:disklifetime} we compare the results of our simulations with available observational estimates based on the disk fractions \citep{Bayo:2012,Ribas:2015}.
We assume that the disk fraction decreases exponentially, following $\exp(-t/t_{\textrm{dis}})$, where $t_{\textrm{dis}}$ is a disk dispersal timescale.
Overall our results show that the disk lifetime is shorter for higher stellar mass,
except in the case with $3\Msun$ which we discuss separately in \secref{sec:4-2}.
 The trend is consistent with the observations.

It is worth noting here that observations toward protoplanetary disks are conducted often in infrared wavelength, and thus thermal emission from dust grains is observed primarily \citep{Ribas:2015,Andrews:2018}.
Both theoretical studies and observations suggest that the gas component and dust are distributed differently \citep{DeGregorioMonsalvo:2013,Ansdell:2018,Toci:2021,Long:2022}.
Recent high-resolution observations have made it possible to observe the gas disk at the $10\au$ scale using a variety of molecular lines \citep{MAPS1}, to show clear images of disk morphology around stars with $1$--$2\Msun$ in the nearby star-forming regions.
We expect that future observations toward a number of disks will reveal evolution of disk morphology around stars with different stellar masses.
In our future work, we compare our simulation results directly with the gas observations to study if a disk loses its mass from outside to inside around low-mass stars and also whether or a disk around a high-mass star has an inner hole at the later stage.

\section{Discussion}
\subsection{disk parameter}
We have performed disk evolution simulations incorporating accretion, MHD winds and photoevaporation with several assumptions and approximations.
For instance, we assume that all the energy liberated by accretion is transferred to drive MHD winds in our fiducial run.  
Here we discuss possible variations of the results due to our choice of various model parameters.

First we study cases when only a part of gravitational energy liberated by accretion is used for winds.
In order to examine the effects quantitatively, we run a series of simulations assuming $10\%$ of liberated gravitational energy is converted to launch MHD winds.

\begin{figure}
    \includegraphics[width=\linewidth,clip]{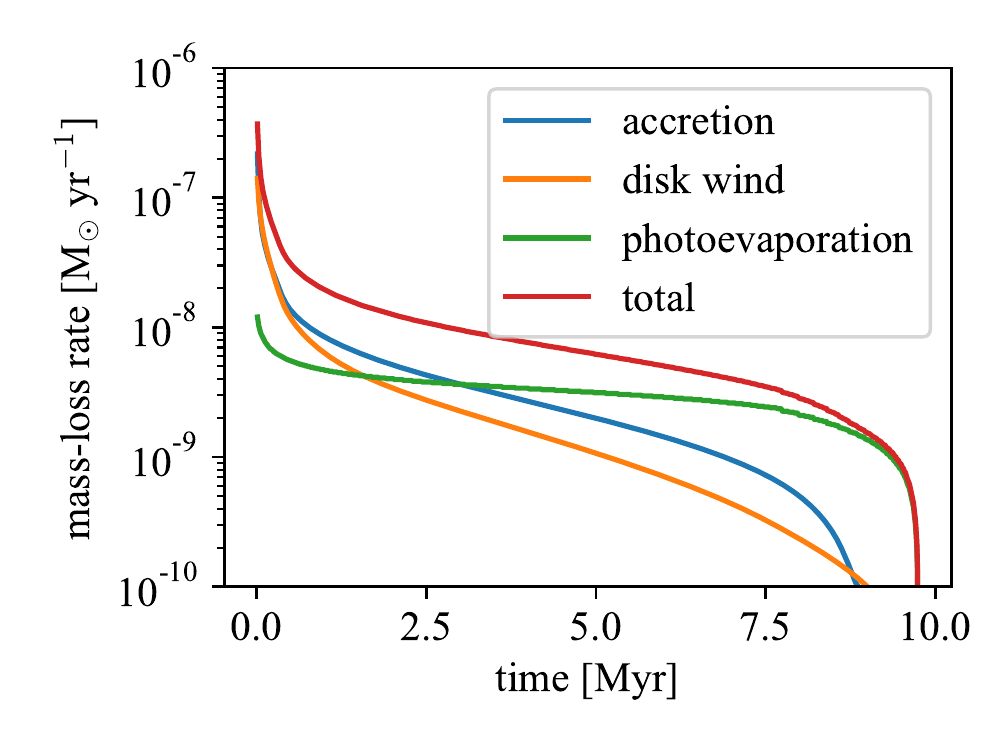}
    \caption{
    Same as \figref{fig:dispersalprocess} but for the weak wind case.
    }
    \label{fig:dispersalprocess-weakwind}
\end{figure}
\figref{fig:dispersalprocess-weakwind} shows the mass-loss rate by each disk dispersal process.
The contribution by MHD winds is lower than in the fiducial case.
At the early phase, accretion and MHD winds are the main dispersal processes.
At the later stage, photoevaporation becomes the dominant process.
In the case of $M_{*}=7\Msun$, the disk is dispersed mainly by photoevaporation.
We estimate the disk lifetime in the same way as the fiducial case.
In the case of $1\Msun$, the disk lifetime is longer than that of the fiducial case only by $\sim1.4\Myr$, which corresponds to 1.2 times of the fiducial case.
Considering the weak wind case is an extreme limit of the reduced mass-loss by MHD winds, we conclude that the mass dependence of the disk lifetime is not affected by the choice of the strong wind or weak wind setups.

\begin{figure}             
    \includegraphics[width=\linewidth,clip]{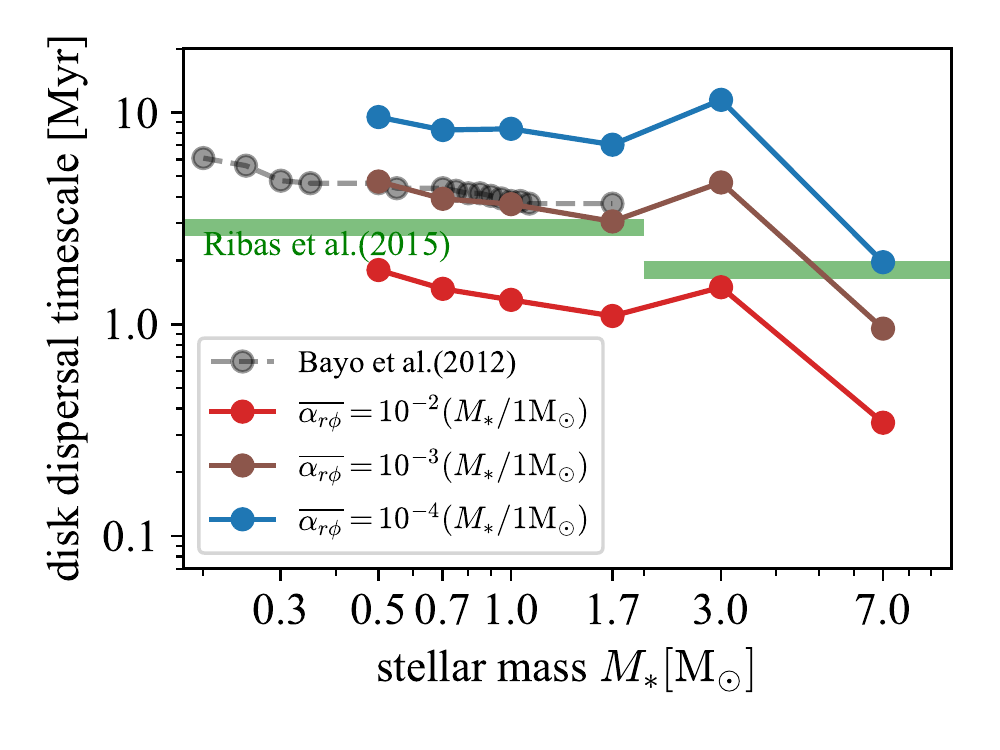}
    \caption{
    The disk dispersal timescale in simulations varying $\overline{\alpha_{r\phi}}$ parameter.
    The blue, brown and red circles are the simulation data with $\overline{\alpha_{r\phi}}=10^{-4}(M_{*}/1\Msun), 10^{-3}(M_{*}/1\Msun)$ and $10^{-2}(M_{*}/1\Msun)$.
    }
    \label{fig:disklifetime-highviscosity}
\end{figure}

Recent observations show that the viscous accretion efficiency, $\alpha$ parameter, varies by a few orders of magnitudes \citep{Hartmann:2016,Ansdell:2018}.
In order to investigate the evolution of disk with high viscous accretion, we also run simulations with $\overline{\alpha}_{r\phi}=10^{-2}(M_{*}/1\Msun)$ and $10^{-3}(M_{*}/1\Msun)$.
Both $\dot{M}_{\textrm{acc}}$ and $\dot{M}_{\textrm{w}}$ increases accordingly.
\figref{fig:disklifetime-highviscosity} shows the resulting disk lifetime with 
different values of $\overline{\alpha}_{r\phi}$.
The lifetime is reduced by 1/5 with an increase in $\overline{\alpha_{r\phi}}$ by a factor of 100.
Interestingly the stellar mass dependence remains essentially the same.
As $\overline{\alpha_{r\phi}}$ increases, both $\dot{M}_{\textrm{acc}}$ and $\dot{M}_{\textrm{w}}$ become larger.
At the same time, $\dot{M}_{\textrm{pe}}$ also increases because the disk radius increases by the strong angular momentum transport caused by the high viscosity.
These effects roughly cancell out, and the overall disk dispersal time is not significantly affected by the choice of $\overline{\alpha_{r\phi}}$.

\subsection{star and disk evolution}
\label{sec:4-2}
\cite{Kunitomo:2021} performed long-term simulations of disk evolution considering accretion, photoevaporation and stellar evolution.
They adopted the photoevaporation mass-loss rate depending on the time varying stellar luminosity in order to study the effect of stellar evolution on disk evolution.
In particular, the X-ray luminosity of a $M_{*}=3\Msun$ star decreases by a factor of $\sim3$ to $t\sim1.5\Myr$.
They clarified that the reduction of X-ray luminosity resulted in the decrease of the mass-loss rate.
They also showed that a disk around a $\sim3\Msun$ star had a long lifetime because of the low X-ray luminosity as we have also found in our simulation.

In our calculations, we use $\dot{\Sigma}_{\textrm{pe}}$ derived from the numerical simulations of \cite{Komaki:2021}.
EUV, FUV and X-ray photons are considered as heating sources, and it is shown that FUV radiation efficiently heats the gas at the disk surface.
According to \cite{Kunitomo:2021}, FUV luminosity increases by at least a few orders of magnitudes for a star with intermediate mass of 1.5--4$\Msun$.
The FUV luminosity increases because the stellar surface becomes hot enough to radiate FUV photons directly.
In the lower-mass side of $M_{*}<2.5\Msun$, the luminosity increases after the age of several $\Myr$, which is comparable to the disk dispersal timescale.
On the other hand, FUV luminosity increases earlier at the age of $\sim1\Myr$ around a $3\Msun$ star.
This is possibly early enough to affect
the disk evolution by photoevaporation.
\cite{Komaki:2021} also run a series of photoevaporation simulations varying the stellar luminosity.
In all the cases of $M_{*}=0.5, 1, 3\Msun$, the mass-loss rates follow the relationship, $\dot{M}_{\textrm{pe}}\propto L_{\textrm{FUV}}^{0.5}$.
In the light of this, we perform a disk evolution calculation around a $3\Msun$ star by incorporating the time-dependent FUV luminosity.
We approximate the evolution of FUV luminosity given by \cite{Kunitomo:2021} with a sigmoid function.
We obtain the surface mass-loss profile by multiplying the same factor as $\dot{M}_{\textrm{pe}}$ from the fiducial value.
The disk around a $3\Msun$ star has a lifetime of $\sim5.1\Myr$, which is half of the fiducial value.
Photoevaporation plays a crucial role in determining the disk lifetime.
It is important to consider evolution of the stellar luminosity around an intermediate-mass star.

\cite{Kunitomo:2021} have shown that the FUV luminosity of a $1\Msun$ star decreases gradually by a factor of 10 from $t=1\Myr$ to $3\Myr$ when the main source of the FUV radiation changes from accretion to radiation from the stellar chromosphere because of the rapid decrease of the accretion rate. 
Since $\dot{M}_{\textrm{pe}}$ decreases as the FUV luminosity decreases \citep{Komaki:2021}, our calculation may overestimate the effect of photoevaporation, and thus may underestimate the disk lifetime.
We expect that the stellar dependence of disk lifetime is steeper if we consider evolution of stellar luminosity.



\subsection{photoevaporation model}
\label{sec:4-3}
We constructed $\dot{\Sigma}_{\textrm{pe}}$ based on \cite{Komaki:2021}.
While previous theoretical studies incorporated EUV and X-ray radiation as a heating source, \cite{Komaki:2021} also considered FUV radiation.
EUV photons are absorbed by the gas near the central star and ionize hydrogen atoms.
As a result, H{\sc ii} regions are formed in the vicinity of the central star, and EUV photons do not reach the outer disk.
FUV photons contribute to the disk heating by photoelectric heating on dust grains.
X-ray photons penetrate into the deeper region of a disk and launch a dense gas flow.
Previous disk evolution simulations take the maximum value among the mass-loss rates by each radiation and incorporate as $\dot{\Sigma}_{\textrm{pe}}$.
The photoevaporation simulations
by \citet{Nakatani:2018a,Nakatani:2018b} and \citet{Komaki:2021} showed
that X-ray photons promoted photoelectric heating by ionization of various species and the decrease of the positive charges of dust grains.
This suggests that it is necessary to perform disk photoevaporation simulations considering FUV and X-ray radiation at the same time.

\begin{figure}
    \includegraphics[width=\linewidth,clip]{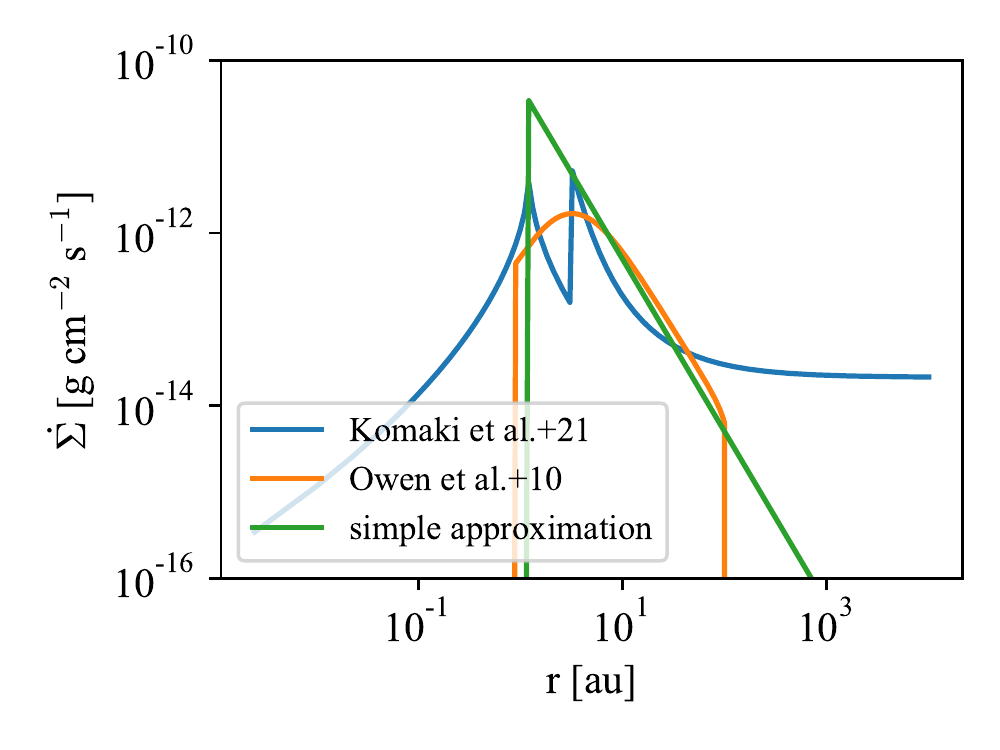}
    \caption{
    The photoevaporation mass-loss profiles: fitted based on \cite{Komaki:2021}(blue), \cite{Owen:2010}(orange) and a simple approximation(green).  
    Since the fitted data by \cite{Owen:2010} is given in an arbitrary unit, the mass-loss profile in the figure is multiplied by $3.1\times10^{-10}$.
    }
    \label{fig:photoevaporationmodel}
\end{figure}

Previous calculations of disk evolution \citep[e.g.,][and more if any]{Kunitomo:2020,Kunitomo:2021} often use $\dot{\Sigma}_{\textrm{pe}}$ given by \cite{Owen:2010}.
\figref{fig:photoevaporationmodel} shows the difference in photoevaporation mass-loss profile by \cite{Owen:2010} and \cite{Komaki:2021} in the case of $M_{*}=1\Msun$.
We multiply by $3.1\times10^{-10}$ 
the profile of \cite{Owen:2010} 
by $3.1\times10^{-10}$ which is given in an arbitrary unit.
We determine the coefficient so that the mass-loss rate inside $100\au$ becomes equal to that of our photoevaporation profile.

The profile of \cite{Komaki:2021} has two peaks generated by EUV and FUV heating, while the profile by \cite{Owen:2010} has one peak.
However, this interesting difference at the inner region of several au does not affect significantly the results of the simulations because MHD winds are the dominant disk dispersal process in the vicinity of the central star.

The computational region of \cite{Komaki:2021} is $[0.89\au, 178\au]$, while that of \cite{Owen:2010} is $[0.82\au, 100\au]$.
In the present paper, we extrapolate the mass-loss profile for the outer disk.
Since photoevaporation becomes the dominant dispersal process at $r>20\au$, the high mass-loss profile results in the fast dispersal in the outer region.
In order to examine the dependence on the mass-loss profile of disk dispersal process, we conduct a disk evolution simulation with a profile given by
\[
\dot{\Sigma}_{\textrm{pe,s}}=3.2\times10^{-12}\gram\cm^{-2}\second^{-1} \left(\frac{L_{\rm{X-ray}}}{10^{30}\erg\second^{-1}}\right)\left(\frac{r}{2.5\au}\right)^{-2},
\]
which is a simple fit for the profile by \cite{Owen:2010}.
This simple profile is shown by an orange line in \figref{fig:photoevaporationmodel}.
Even though \cite{Owen:2010} only provides a profile within $100\au$, we extend the fit to $r>100\au$ so that we incorporate photoevaporation in the outer disk.
We set the inner boundary to $0.14\rg=1.2\au$ following \cite{Liffman:2003}, which calculate the gravitational radius theoretically.
The dominant disk dispersal process changes from MHD winds to photoevaporation at the age of $\sim2\Myr$.
\begin{figure*}
    \includegraphics[width=\linewidth,clip]{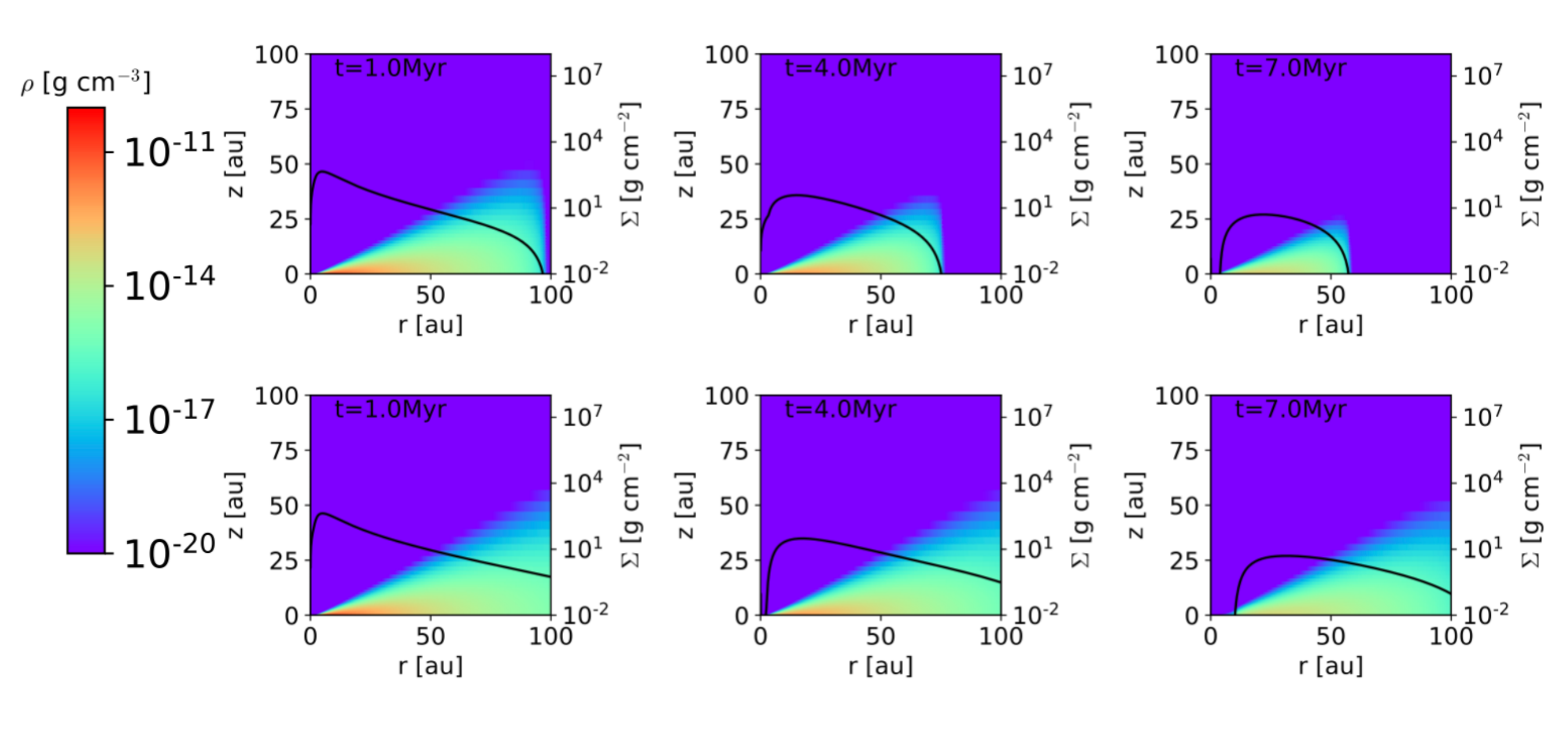}
    \caption{
    Snapshots of the reconstructed gas density in a simulation with $M_{*}=1\Msun$ at $t=1, 4, 7\Myr$.
    The color map shows the density and the black line expresses the surface density.
    The top row shows the density distribution in the fiducial simulation and the bottom row shows the density distribution in the simulation with the simple photoevaporation profile                                       .
    }
    \label{fig:snapshots-simplephotoevaporation}
\end{figure*}
\figref{fig:snapshots-simplephotoevaporation} shows the reconstructed 2D snapshots of the simulation.
As shown in \figref{fig:photoevaporationmodel}, $\dot{\Sigma}_{\textrm{pe,s}}$ decreases with increasing distance from the central star following $r^{-2}$, while the profile by \cite{Komaki:2021} maintains $\sim3.0\times10^{-14}\gram\cm^{-2}\second^{-1}$ in the outer region.
The disk at a several tens of $\au$ remains at the last stage of disk dissipation.
Following these, we need to apply realistic photoevaporation profile to obtain the disk density distribution.

\subsection{Other effects}
We assume $\dot{M}_{\textrm{pe}}$ is constant throughout our calculations.
It has been suggested that photoevaporation can be enhanced when an inner hole opens \citep{Alexander:2014, Owen:2010, Nakatani:2021}. 
In our study, only the simulation with $M_{*}=7\Msun$ shows the formation of a hole at $\sim1\Myr$, before the disk disperses. The time evolution of $\dot{M}_{\textrm{pe}}$, especially when a hole or a gap is formed, may need to be taken into account to
evaluate the disk lifetime more accurately.

Disk observations have shown that disks in a strong radiation field have specific shapes with long tails  \citep{Odell:1993,McCaughreanOdell:1996,Winter:2018}.
Theoretical studies suggested that the dense gas was eroded by high radiation from the nearby high-mass stars \citep{RichlingYorke:1998,Haworth:2019}.
\cite{Haworth:2019} performed 2D hydrodynamics simulations.
They assumed that the disk was exposed to the high FUV radiation fields and solved radiation hydrodynamics to calculate the mass-loss rate.
As s result, the outer disk is effectively removed because the thin disk is heated efficiently by the radiation from the nearby stars.
The present study does not consider radiation by nearby stars.
We expect that the outer disk would be dispersed more efficiently under an external radiation field, and the disk radius would shrink rapidly.

\section{Summary}
Recent PPD observations in star forming regions suggested that the lifetime decreases with increasing central stellar mass.
Theoretical studies proposed three disk dispersal mechanisms: accretion, MHD winds and photoevaporation.
In order to understand both the typical lifetime of a few million years and the stellar mass dependence, it is necessary to conduct disk evolution simulations throughout the disk lifetime considering all the disk dispersal mechanisms.
We performed 1D disk evolution simulations varying the central stellar mass in the range of 0.5--7$\Msun$.
We showed the disk loses its mass mainly by MHD winds in the early stage, and later by photoevaporation.
Especially, in the case with $M_{*}=7\Msun$, photoevaporation by the high radiation contributes more than other cases, and a gap opens at $\sim1\Myr$.
We find that photoevaporation shapes the disk morphology by clearing out the gas in the outer disk efficiently.
The time evolution of disk mass can be described by a simple function that is given by transforming the sigmoid function,
which yields an accurate estimate of the disk dispersal time.
The dispersal time around a high-mass star is $\sim2\Myr$, which is shorter than the other cases by the factor of $\sim5$.
The trend is consistent with recent observations.
Finally, the mass dependence of the dispersal timescale does not vary by the choice of strong wind or weak wind or the choice of viscous parameter.

\acknowledgments
This research was supported by Forefront Physics and Mathematics Program to Drive Transformation (FoPM), a World-leading Innovative Graduate Study (WINGS) Program, the University of Tokyo.
NY acknowledges financial support
from MEXT/JSPS KAKENHI 20H05844.
T.K.S. is supported in part by Grants-in-Aid for Scientific Research from the MEXT/JSPS of Japan, 17H01105, 21H00033 and 22H01263 and by Program for Promoting Research on the Supercomputer Fugaku by the RIKEN Center for Computational Science (Toward a unified view of the universe: from large-scale structures to planets; grant 20351188-PI J. Makino) from the MEXT of Japan.

%
\vspace{5mm}






\bibliography{bibliography}
\bibliographystyle{aasjournal}



\end{document}